\documentclass{amsart}
\usepackage[letterpaper,hmargin=1in,vmargin=1.25in]{geometry}

\usepackage{amsmath, amsthm, amsfonts, amssymb}
\newtheorem{theorem}{Theorem}[section]
\newtheorem{lemma}[theorem]{Lemma}

\usepackage{tikz}
\usetikzlibrary{matrix,arrows}

\theoremstyle{definition}
\newtheorem{definition}[theorem]{Definition}

\newtheorem{heuristic}[theorem]{Heuristic Assumption}

\theoremstyle{remark}
\newtheorem{remark}[theorem]{Remark}

\numberwithin{equation}{section}

\makeatletter
\providecommand*{\twoheadrightarrowfill@}{%
  \arrowfill@\relbar\relbar\twoheadrightarrow
}
\providecommand*{\twoheadleftarrowfill@}{%
  \arrowfill@\twoheadleftarrow\relbar\relbar
}
\providecommand*{\xtwoheadrightarrow}[2][]{%
  \ext@arrow 0579\twoheadrightarrowfill@{#1}{#2}%
}
\providecommand*{\xtwoheadleftarrow}[2][]{%
  \ext@arrow 5097\twoheadleftarrowfill@{#1}{#2}%
}
\makeatother

\makeatletter
\providecommand*{\hookrightarrowfill@}{%
  \arrowfill@\relbar\relbar\hookrightarrow
}
\providecommand*{\hookleftarrowfill@}{%
  \arrowfill@\twoheadleftarrow\relbar\relbar
}
\providecommand*{\xhookrightarrow}[2][]{%
  \ext@arrow 0579\hookrightarrowfill@{#1}{#2}%
}
\providecommand*{\hookleftarrow}[2][]{%
  \ext@arrow 5097\hookleftarrowfill@{#1}{#2}%
}
\makeatother

\newcommand{\Z}{\mathbb Z}
\newcommand{\F}{\mathbb F}

\begin{document}
\title[Computing discrete logarithms in subfields of residue class rings.]{Computing discrete logarithms in subfields of residue class rings.}

\author{Ming-Deh Huang }
\address{Computer Science Department, University of Southern California, Los Angeles.}
\email{mdhuang@usc.edu }

\author{Anand Kumar Narayanan}
\address{Computer Science Department, University of Southern California, Los Angeles.}
\email{aknaraya@usc.edu}

\begin{abstract}
Recent breakthrough methods  \cite{gggz,joux,bgjt} on computing discrete logarithms in small characteristic finite fields share an interesting feature in common with the earlier medium prime function field sieve method \cite{jl}. To solve discrete logarithms in a finite extension of a finite field $\F$, a polynomial $h(x) \in \F[x]$ of a special form is constructed with an irreducible factor $g(x) \in \F[x]$ of the desired degree. The special form of $h(x)$ is then exploited in generating multiplicative relations that hold in the residue class ring $\F[x]/h(x)\F[x]$ hence also in the target residue class field $\F[x]/g(x)\F[x]$.
An interesting question in this context and addressed in this paper is: when and how does a set of relations on the residue class ring determine the discrete logarithms in the finite fields contained in it? We give necessary and sufficient conditions for a set of relations on the residue class ring to determine discrete logarithms in the finite fields contained in it. We also present efficient algorithms to derive discrete logarithms from the relations when the conditions are met.
The derived necessary conditions allow us to clearly identify structural obstructions intrinsic to the special polynomial $h(x)$ in each of the aforementioned methods, and propose modifications to the selection of $h(x)$ so as to avoid obstructions.
Joux's relation generation algorithm with our modified polynomial selection, the Barbulescu-Gaudry-Joux-Thome descent, the Pohlig-Hellman algorithm and our method for determining discrete logarithms in subfields of residue class rings together imply a heuristic quasi polynomial time algorithm for computing discrete logarithms in small characteristic finite fields.  In addition, a generator (primitive element) for the multiplicative group of the finite field can be efficiently constructed by our method.  This is especially interesting when the factorization of the order of the unit group $(\F[x]/g(x)\F[x])^\times$ is not known.
\end{abstract}
\maketitle

\section{Introduction}\label{intro}
\ \\ The discrete logarithm problem over a finite cyclic group $G$ is given a generator $\eta \in G$ and an element $\gamma \in G$ to find an integer $\log_\eta(\gamma)$ uniquely determined modulo $|G|$ such that $\gamma = \eta ^{\log_\eta(\gamma)}$. The discrete logarithm problem over the multiplicative group $\F_{p^n}^\times$ of the finite field $\F_{p^n}$ serves as an important cryptographic primitive. For instance, the security of the Diffie-Hellman \cite{dh} key exchange protocol and ElGamal's cryptosystem \cite{elg} are conditioned on the hardness of the discrete logarithm problem over finite fields.\\ \\
Let $L(\ell)$ denote $\exp( \mathcal{O}((\log(p^n)^\ell)(\log\log(p^n))^{1-\ell}))$. The index calculus method has been developed for the discrete logarithm problem over finite fields in a series of works \cite{adl}\cite{cop}\cite{gor}\cite{adl1}\cite{ah}\cite{jl}\cite{jlsv} establishing that the problem can be solved in $L(1/3)$ time which is subexponential in $\log(p^n)$.\\ \\
In recent breakthroughs, Gologlu, Granger, McGuire, Zumbragel \cite{gggz} and Joux \cite{joux} independently devised algorithms that assuming certain heuristics compute discrete logarithms in small characteristic finite fields faster than previously known.
In a recent further advancement Barbulescu, Gaudry, Joux and Thome \cite{bgjt} proposed an algorithm for computing discrete logarithms in small characteristic finite fields in quasi polynomial time. The finite field representation chosen in \cite{bgjt} is identical to the one in Joux's algorithm \cite{joux}. The descent phase in \cite{bgjt} expresses a power of an arbitrary element in the multiplicative group of the finite field as a product of powers of elements in the factorbase in \cite{joux}. Thereby, the descent in \cite{bgjt} reduces the discrete logarithm computation over the finite field to computing discrete logarithms between elements in the factorbase which can be solved by the initial phase in Joux's algorithm \cite{joux}.\\ \\
The aforementioned methods \cite{gggz,joux,bgjt}, as well as the earlier medium prime function field sieve method \cite{jl}, share the following interesting feature. To solve discrete logarithms in a finite extension of a finite field $\F$, a polynomial $h(x) \in \F[x]$ of a special form is constructed with an irreducible factor $g(x) \in \F[x]$ of the desired degree. The special form of $h(x)$ is exploited in generating multiplicative relations and 
heuristic assumptions suggest that once enough relations are  collected, standard linear algebraic methods yield discrete logarithms in $(\F[x]/g(x)\F[x])^\times$.
However obstructions can be identified that prevent these heuristics from being true in general. These obstruction arise as a consequence of the fact that the collected relations from which discrete logarithms are to be derived hold not only in the target residue class field $\F[x]/g(x)\F[x]$ determined by $g(x)$ but also in a larger structure, the residue class ring $\F[x]/h(x)\F[x]$ determined by the special polynomial $h(x)$. The nature of the obstruction is structural and intrinsic to $h(x)$ and would arise regardless of how many relations are generated.\\ \\
Even in cases where obstructions do not arise, there are issues in applying the standard algebraic methods to the collected relations (see remark \ref{rem}). One issue is that the factorbase may not contain a generator for $(\F[x]/g(x)\F[x])^\times$. Another issue is that the factorization of $|(\F[x]/g(x)\F[x])^\times|$ may not be known. In typical implementations of the Diffie-Hellman protocol over finite fields \cite{dh}, the field size is chosen such that the order of the multiplicative group has a single large prime factor and hence the factorization of the group order is known. However, there are cases of cryptographic significance wherein the group order is not known. For instance, the MOV-attack \cite{mov} reduces the elliptic curve discrete logarithm problem to a computing discrete logarithm over a finite extension of a finite field where in general the factorization of the order of the multiplicative group is not known.\\ \\
These issues raise the following interesting question. When and how does a set of relations on the residue class ring determine the discrete logarithms in the finite fields contained in it? We address this question in \S~\ref{residue_class_rings} where in we describe necessary and sufficient conditions for a set of relations on the residue class ring to determine discrete logarithms in the finite fields contained in it. Further, we present efficient algorithms to derive discrete logarithms from the relations when these conditions are met.  The derived necessary conditions allow us to clearly identify structural obstructions intrinsic to the special polynomial $h(x)$ in each of the aforementioned methods, and propose modifications to the selection of $h(x)$ so as to avoid obstructions.\\ \\
The issue that the relations generated in Joux's algorithm hold in $\F[x]/h(x)\F[x]$ and hence the need for modification was raised independently by Cheng, Wan and Zhang \cite{cwz} and Huang and Narayanan \cite{primitive}.\\ \\
We remark that these obstructions do not arise (see \S~\ref{obstruction_joux}) in the case when the degree $\deg(g)$ of the field extension considered equals $|\F^\times|$. This is often referred to as the Kummer case and was 
the setting for some of the experimental results in favor of the heuristics implicit in \cite{joux}.

\ \\In addition Cheng, Wan and Zhang \cite{cwz} described potential traps to the descent algorithm in \cite{bgjt} that prevent the descent from succeeding and suggested a trap avoiding descent. In the trap avoiding descent of \cite{cwz}, certain relations involving the factors of $h(x)$ are identified as traps and are not used when encountered. This trap avoidance thus comes at the cost of discarding relations. In \S~\ref{obstruction_descent}, we observe that some of the relations dropped in fear of traps could be salvaged. Further, these salvaged relations serve in further breaking the symmetry between $g(x)$ and the other irreducible factors of $h(x)$.

\ \\Joux's relation generation algorithm with our modified polynomial selection, the Barbulescu-Gaudry-Joux-Thome descent, the Pohlig-Hellman algorithm and the algorithms in \S~\ref{residue_class_rings} for determining discrete logarithms in subfields of residue class rings together imply a heuristic quasi polynomial time algorithm for computing discrete logarithms in small characteristic finite fields. In addition, a generator (primitive element) for the multiplicative group of the finite field can be efficiently constructed by our method.  This is especially interesting when the factorization of the order of the multiplicative group is not known.

\ \\ The current paper subsumes an earlier account of these results we posted at \cite{hn2}.  
\subsection{Summary of Results}\label{summary}
\ \\Algorithms for computing discrete logarithms over finite fields typically involve two steps. First a subset of the multiplicative group often referred to as the factorbase is chosen and the discrete logarithms of the elements in the factorbase are determined by collecting multiplicative relations between the elements in the factorbase. Then the element whose discrete logarithm is sought is written as a product of powers of elements in the factorbase thereby determining its discrete logarithm.\\ \\
For a non constant $f(x) \in \F[x]$, let $\F_f$ denote the residue class ring $\F[x]/\left( f(x)\F[x]\right)$, and let $\F_f^{\times}$ denote the subgroup of units in $\F_f$ (the multiplicative subgroup of $\F_f$).
Let $S \subset \F[x]$ be a finite set such that none of the elements in $S$ share a non trivial factor with $h(x)$. Let $\Z S$ be the free $\Z$-module generated by $S$.  Let $R$ be a finite subset of $\Z S$ and let $\Z R$ be the submodule of $\Z S$ generated by $R$.  Then $\Z R$ is the set of relations for the (additive) group $\Z S/\Z R$.
For a non constant $f(x) \in \F[x]$, let $S_f$ denote $\{ d(x) \mod f(x) | d(x) \in S\}$ and call $\sum_{s \in S} e_s s \in \Z S$ as a relation for $\langle S_f \rangle \subset \F_f^{\times}$ if  and only if $$\prod_{s \in S} s(x)^{e_s} = 1 \mod f(x).$$
Let $h(x) \in \F[x]$ be a polynomial, and let $g(x) \in \F[x]$ be an irreducible factor of $h(x)$.
Since $g(x)$ divides $h(x)$, a relation for $\langle S_h \rangle$ is also a relation for $\langle S_g \rangle$.\\ \\
Consider an algorithm that attempts to solve discrete logarithms in $\F_g^\times$ of elements in $S_g$ by collecting a finite set $R \subset \Z S$ of relations for $\langle S_h \rangle$.\\ \\
Since $\F_g^\times$ is cyclic, $\langle S_g \rangle$ is cyclic and we may pick a $\gamma$ such that $\langle \gamma \rangle = \langle S_g \rangle$. Once such a $\gamma$ is chosen, $ \forall s_i \in S$ there exists $a_i \in \Z$ determined uniquely modulo $|\F_g^\times|$ such that $\gamma^{a_i} = s_i$. In particular, $\forall s_i,s_j \in S$, $s_i^{a_j}=s_j^{a_i}$ and we have the relation $r_{i,j} := a_j s_i - a_i s_j$ for $\langle S_g \rangle$. The following definitions capture what it means for discrete logarithms to be determined from a set of relations by linear algebraic means.  
\begin{definition}\label{dlog_determine}
Discrete logarithms in $S_g$ are said to be determined by $R$ if $\forall s_i,s_j \in S, r_{i,j} \in \Z R$.
\end{definition}
\begin{definition}\label{dlog_determine_modL}
For a positive integer $L$, discrete logarithms in $S_g$ are said to be determined modulo $L$ by $R$ if $\forall s_i,s_j \in S, r_{i,j} \mod L \in \Z R \otimes \Z/L\Z$.
\end{definition}
\ \\Since the vector $(a_s)_{s \in S}$ is uniquely determined up to a $\Z$ multiple that depends on the choice of the generator $\gamma$, both definitions do not depend on $\gamma$.
\\ \\We show that for prime $\ell$ such that the subgroup generated by $S_g$ has nontrivial $\ell$-part, discrete logarithms in $S_g$ are determined modulo $\ell$ by $R$ if and only if the $\Z/\ell\Z$ rank of $\Z R \otimes \Z/\ell\Z$ is $|S|-1$.
For a positive integer $L$ dividing the order of $\F_g^{\times}$ such that $L$ is prime to $|\F_g^{\times}|/L$, when the rank condition is met for all primes $\ell | L$, then one can efficiently construct a generator for $\F_g^{\times} [L]$ and determine the discrete logarithms of all elements of $S$ modulo $L$.  On the other hand the $\Z/\ell\Z$ rank of $\Z R \otimes \Z/\ell\Z$ is $|S|-1$
 if and only if $(\Z S/\Z R)/\ell(\Z S/\Z R)$ is cyclic and non-trivial, and from this one can clearly identify obstructions for $R$ to determine discrete logarithms modulo $\ell$ for $\F_g^{\times}$. More precisely we have the following.
 \begin{theorem}\label{theorem_factorbase}\ \\
 \begin{enumerate}
 \item For prime $\ell$ such that the subgroup generated by $S_g$ has nontrivial $\ell$-part, discrete logarithms in $S_g$ are determined modulo $\ell$ by $R$ if and only if the $\Z/\ell\Z$ rank of $\Z R \otimes \Z/\ell\Z$ is $|S|-1$  if and only if $(\Z S/\Z R)/\ell(\Z S/\Z R)$ is cyclic and non-trivial.
 \item For positive integer $L$ dividing the order of $\F_g^{\times}$ such that $L$ is prime to $|\F_g^{\times}|/L$, if the $\Z/\ell\Z$ rank of $\Z R \otimes \Z/\ell\Z$ is $|S|-1$ for all primes $\ell | L$, then in time polynomial in $\log(L)$, $|S|$ and $|R|$ we can construct a generator for $\F_g^{\times} [L]$ and determine the discrete logarithms of the projection to $\F_g^{\times} [L]$ of every element of $S$ with respect to the generator.
 \item For prime $\ell$ dividing the order of $\F_g^{\times}$, if $\langle S_h \rangle /\langle S_h \rangle^{\ell}$ is not cyclic then discrete logarithms of $S_g$ modulo $\ell$ cannot be determined from $R$.
\end{enumerate}
 \end{theorem}
\ \\Part (1) and (2) of theorem \ref{theorem_factorbase} follow from lemma \ref{necessary} and theorem \ref{theorem_g} of \S~\ref{dlog_factorbase}.\\ \\
Part (3) of the theorem follows from Part (1) and the natural surjection $(\Z S/\Z R)/\ell(\Z S/\Z R) \twoheadrightarrow \langle S_h \rangle /\langle S_h \rangle^{\ell}$.\\ \\
Further, in \S~\ref{obstruction_joux} we argue that if there is a prime $\ell$ dividing $|\F_h^\times|$ such that $\F_h^\times/(\F_h^\times)^\ell$ is not cyclic, then $\langle S_h \rangle/\langle S_h \rangle^\ell$ is not likely to be cyclic thereby resulting in an obstruction.

\ \\ We next turn to the descent phase. Let $v \in \F[x]$ and let $U \subset \F[x]$ be a finite set (It may be helpful to think of $U$ as the factorbase $S$). We say that $v$ is descent to $U$ through a set of relations $T$ for $\F_h^{\times}$ if there exists $a_u \in \Z$ such that $v - \sum_{u \in U}{a_u u} \in \Z T$.\\ \\
More generally, for a finite set $V \in \F[x]$ disjoint from $U$, let $T \subset \Z (U \cup V)$ be a set of relations for the subgroup of $\F_h^\times$ generated by $U \cup V$ modulo $h$. For $r=\sum_{v \in V}b_v v + \sum_{u \in U} b_u u \in T$, let $r_V = \sum_{v \in V}b_v v$. Let $T_V$ denote $\{r_V| r \in T\}$. We say that $V$ is descent to $U$ through $T$ if every $v\in V$ is descent to $U$ through $T$.  From the definition it follows that $V$ is descent to $U$ through $T$ if and only if $\Z T_V = \Z V$. Likewise for a positive integer $L$, one can define what it means for $V$ to be descent to $U$ through $T$ modulo $L$ in terms of $\F_h^\times/(\F_h^\times)^{L}$.  It is then clear that the descent is possible if and only if $\Z T_V  \otimes \Z/L\Z= \Z V \otimes \Z/L\Z$.

\ \\ In \S~\ref{algorithm_descent}, we show that for prime $\ell$ such that the subgroup generated by $S_g$ has nontrivial $\ell$-part, descent from $V$ to $U$ modulo $\ell$ through $T$ is possible if and only if the $\Z/\ell\Z$ rank of $\Z T_V  \otimes \Z/\ell\Z$ is $|V|$. For a positive integer $L$ dividing the order of $\F_g^{\times}$ such that $L$ is prime to $|\F_g^{\times}|/L$, when the rank condition is met for all primes $\ell | L$, then one can efficiently descend from $V$ to $U$ through $T$ modulo $L$. On the other hand the $\Z/\ell\Z$ rank of $T_V \otimes \Z/\ell\Z$ is $|V|$ if and only if $(\Z S/\Z T_V)/\ell(\Z S/\Z T_V)$ is trivial, and from this one can clearly identify obstructions for descent from $U$ to $V$ through $T$ modulo $\ell$. More precisely we have the following.
\begin{theorem}\label{theorem_descent}\ \\
\begin{enumerate}
\item \label{necessary_descent} For a prime $\ell$, descent from $V$ to $U$ through $T$ modulo $\ell$ is possible if and only if the $\Z/\ell\Z$ rank of $\Z T_V  \otimes \Z/\ell\Z$ is $|V|$ if and only if $(\Z S/\Z T_V)/\ell(\Z S/\Z T_V)$ is trivial.
\item Let $L$ be a positive integer such that $L \mid |\F_g^\times|$ and $\gcd(L,|\F_g^\times|/L)=1$. If for all prime $\ell$ dividing $L$, the $\Z/\ell\Z$ rank of $\Z T_V \otimes \Z/\ell\Z$ is $|V|$, in time polynomial in $\log(L)$, $|S|$ and $|T|$, we can find for every $v \in V$, $w \in \Z$ and $w_u \in \Z, \forall u\in U$ such that the projections of $v(x)^{w}$ and $\prod_{u \in U} u(x)^{w_u} $ to $\F_h^\times[L]$ are identical.
\item If there is a prime $\ell$ dividing $|\F_g^\times|$ such that the image of $U$ in $(\F_h^\times)/(\F_h^\times)^\ell$ generates a proper subgroup of $(\F_h^\times)/(\F_h^\times)^\ell$, then heuristically the descent to $U$ using relations for $\F_h^\times$ modulo $\ell$ fails with probability at least $1- 1/\ell$.
\end{enumerate}
\end{theorem}
\ \\Part (1) and (2) of theorem \ref{theorem_descent} follow from lemma \ref{descent_necessary} and theorem \ref{theorem_gg} of \S~\ref{algorithm_descent}.\\ \\
Part (3) of the theorem follows from (1), since if the image of $U$ in $(\F_h^\times)/(\F_h^\times)^\ell$ generates a proper subgroup of $(\F_h^\times)/(\F_h^\times)^\ell$, then the cardinality of the subgroup is at most $1/\ell$ times $|(\F_h^\times)/(\F_h^\times)^\ell|$.\\ \\
In particular, consider the last step in the descent where we attempt to descend to the factorbase $S$, that is $U = S$. Let $G_U$ denote the subgroup of $(\F_h^\times)/(\F_h^\times)^\ell$ generated by the image of $U$ in $(\F_h^\times)/(\F_h^\times)^\ell$. If $(\F_h^\times)/(\F_h^\times)^\ell$ is not cyclic, then $G_U$ is either a proper subgroup of $(\F_h^\times)/(\F_h^\times)^\ell$ or $G_U = (\F_h^\times)/(\F_h^\times)^\ell$. In the former case, the obstruction in part $(3)$ of theorem \ref{theorem_descent} occurs and in the latter case, since $\langle S_h \rangle / \langle S_h \rangle^\ell$ is not cyclic, the obstruction in part (3) of theorem \ref{theorem_factorbase} occurs and we have the following theorem.
\begin{theorem}\label{obstruction_theorem}
\textit{For algorithms restricted to generating relations for $\F_h^\times$, if there is a prime $\ell$ dividing $|\F_g^\times|$ such that the $\ell$ primary part of $\F_h^\times$ is not cyclic, then either the determination of discrete logarithms in the factorbase modulo $\ell$ fails or the last step in the descent is to likely fail modulo $\ell$.}
\end{theorem}
\ \\ An immediate consequence of theorem \ref{obstruction_theorem} is that unless $\deg(h)=\deg(g)$, $\F^\times \times \F^\times$ is contained in $\F_h^\times$ and hence there is an obstruction for every prime $\ell$ dividing $|\F^\times|$. However, in the context of the algorithms of \cite{joux,bgjt}, $|\F^\times|$ is small compared to the group order $|\F^{\deg(g)}|$. By setting a smoothness bound greater than $|\F|$, the discrete logarithms modulo $\ell$ dividing $|\F^\times|$ may be computed using the Pohlig-Hellman \cite{ph} algorithm. Following theorems \ref{theorem_factorbase} and \ref{obstruction_theorem}, our strategy is to consider $|\F_g^\times|$ as a product of a smooth factor and a non smooth factor $L$. In \S~\ref{poly_select}, we impose conditions on $h(x)$ that ensure that the obstructions in theorems \ref{theorem_factorbase}(3) and \ref{theorem_descent}(3) do not occur for primes dividing $L$. In particular, requiring that $g^2 \nmid h$ and that $\gcd(|\F_g^\times|,|\F_{h/g}^\times|)$ is smooth, ensure that for every prime $\ell$ dividing $L$, the $\ell$ primary part of $\F_h^\times$ is cyclic thereby avoiding the obstructions.\\ \\
In \S~\ref{obstruction_joux} we apply theorem \ref{theorem_factorbase} to Joux's algorithm. In particular, we propose a modified polynomial selection (see \S~\ref{obstruction_joux}) that picks a $h(x)$ that avoids the obstructions for both the Joux's algorithm and the Barbulescu-Gaudry-Joux-Thome descent.\\ \\
In \S~\ref{obstruction_descent}, we present a refinement of the Barbulescu-Gaudry-Joux-Thome descent to tackle the issue of traps. Theorem \ref{theorem_descent} is then applied to the refined descent algorithm.\\ \\ 
With the restrictions placed on $h(x)$, heuristically the relations generated by the algorithms of Joux and the Barbulescu-Gaudry-Joux-Thome descent with refinement seem adequate to satisfy the rank conditions in theorems \ref{theorem_factorbase}(2) and \ref{theorem_descent}(2) respectively. If they indeed satisfy the rank conditions, then by theorems \ref{theorem_factorbase} (2) and \ref{theorem_descent} (2) it follows that we can find a generator and with respect to it solve discrete logarithms in the non-smooth component $\F_g^\times[L]$. The discrete logarithm computation in the smooth component can performed efficiently by the Pohlig-Hellman algorithm.\\ \\
Theorems \ref{dlog_factorbase_theorem} and \ref{dlog_descent_theorem} imply the following theorem.
\begin{theorem}
Under heuristic assumptions \ref{heu1}, \ref{heuristic_factorbase} and \ref{heuristic_descent}, there is a deterministic algorithm that finds a generator for $\F_{p^n}^\times$ in time polynomial in $p$ and $n$ and finds discrete logarithms with respect to that generator in time polynomial in $p^{\log(n)}$ and $n^{\log(n)}$.
\end{theorem}

\section{Computing Discrete Logarithms in Residue Class Rings}\label{residue_class_rings}
\subsection{Discrete Logarithms in the Factorbase}\label{dlog_factorbase}
We keep the notation from \S~\ref{summary}. For a finite group $G$, a prime $\ell$ and a positive integer $a$, let $G(\ell)$ denote the $\ell$-primary part of $G$ and $G[a]$ the $a$-torsion.
\begin{lemma}\label{necessary}
Let $\ell$ be a prime dividing $|\F_g^\times|$. If $\langle S_g \rangle(\ell) = \F_g^\times(\ell)$, then $R$ determines discrete logarithms in $S_g$ modulo $\ell$ if and only if the $\Z/\ell\Z$ rank of $\Z R \otimes \Z/\ell\Z$ is $|S|-1$ if and only if $(\Z S/ \Z R)/\ell(\Z S/\Z R)$ is cyclic and non-trivial.
\end{lemma}

\ \\ Tensoring with $\Z/\ell\Z$, the exact sequence $$0 \longrightarrow  \Z R  \longrightarrow \Z S \longrightarrow \Z S/\Z R \longrightarrow 0$$
induces the sequence 
$$  \Z R \otimes \Z/\ell\Z \longrightarrow \Z S \otimes \Z/\ell\Z \longrightarrow (\Z S/\Z R) \otimes \Z/\ell\Z \longrightarrow 0.$$
which is exact due to the right exactness of tensoring. Thus we have, $$(\Z S \otimes \Z/\ell\Z)/(\Z R \otimes \Z/\ell\Z) \cong (\Z S /\Z R)/\ell(\Z S/\Z R).$$
Both $(\Z S \otimes \Z/\ell\Z)$ and $(\Z R \otimes \Z/\ell\Z)$ are $\Z/\ell\Z$ vector spaces and $(\Z S \otimes \Z/\ell\Z)$ is $|S|$ dimensional over $\Z/\ell\Z$.
We thus have the following characterization.
\begin{enumerate}
\item $(\Z S/\Z R)/\ell(\Z S/\Z R)$ is trivial $\Leftrightarrow$\  $\Z/\ell\Z$ rank of $\Z R \otimes \Z/\ell\Z$ is $|S|$.
\item $(\Z S/\Z R)/\ell(\Z S/\Z R)$ is cyclic and non-trivial $\Leftrightarrow$\ $\Z/\ell\Z$ rank of $\Z R \otimes \Z/\ell\Z$ is $|S|-1$.
\item $(\Z S/\Z R)/\ell(\Z S/\Z R)$ is not cyclic $\Leftrightarrow$\ $\Z/\ell\Z$ rank of $\Z R \otimes \Z/\ell\Z$ less than $|S|-1$.\\
\end{enumerate}
Since $\Z R$ is a subset of the relations for $\langle S_g \rangle$, there is a surjection $$\Z S/\Z R \twoheadrightarrow \langle S_g \rangle$$ which implies that there is a surjection $$\phi_\ell: (\Z S/\Z R)/\ell(\Z S/\Z R) \twoheadrightarrow  \langle S_g \rangle/ \langle S_g\rangle^\ell.$$
Since $\langle S_g \rangle(\ell) = \F_g^\times(\ell)$, $\langle S_g \rangle/ \langle S_g\rangle^\ell \cong \F_g^\times/(\F_g^\times)^\ell$. Further, since $\ell$ divides $|\F_g^\times|$, $\langle S_g \rangle/ \langle S_g\rangle^\ell$ is not trivial. Thus $(\Z S/\Z R)/\ell(\Z S/\Z R)$ is not trivial and the $\Z/\ell\Z$ rank of $\Z R \otimes \Z/\ell\Z$ is at most $|S|-1$.\\ \\
For $(\Z S \otimes \Z/\ell\Z)/(\Z R \otimes \Z/\ell\Z) (\cong (\Z S/\Z R)/\ell(\Z S/\Z R))$ to determine discrete logarithms in $S_g$ modulo $\ell$, it is necessary that $(\Z S/\Z R)/\ell(\Z S/\Z R)$ is cyclic.\\ \\
If the $\Z/\ell\Z$ rank of $\Z R \otimes \Z/\ell\Z$ is $|S|-1$, then discrete logarithm in $(\Z S \otimes \Z/\ell\Z)/(\Z R \otimes \Z/\ell\Z) (\cong (\Z S/\Z R)/\ell(\Z S/\Z R))$ of the images of the basis elements can be determined by solving a linear system. Thereby under $\phi_\ell$, the discrete logarithms in $\langle S_g \rangle/ \langle S_g\rangle^\ell (\cong\F_g^\times/(\F_g^\times)^\ell)$ of the images of elements in $S_g$ is determined. $\square$\\ \\
Let $M$ denote the matrix whose rows consist of the relation vectors corresponding to the set of collected relations $R$. For a positive integer $L$, Let $M_L$ denote the matrix with entries of $M$ reduced modulo $L$. By definition, the $\Z/L\Z$ module generated by the rows of $M_L$ is $\Z R \otimes \Z/L\Z$.
\begin{lemma}\label{gaussian} Let $L$ be a positive integer. If for all prime $\ell$ dividing $L$, the $\Z/\ell\Z$ rank of $M_\ell$ is $|S|-1$, then in time polynomial in $\log(L)$, $|R|$ and $|S|$, we can find a generator of $(\Z S/\Z R)/L(\Z S/\Z R)$ and for every $s \in S$ compute the discrete logarithm of its image in $(\Z S/\Z R)/L(\Z S/\Z R)$ with respect to the found generator. 
\end{lemma}
\ \\ We first show that if for all prime $\ell$ dividing $L$, the $\Z/\ell\Z$ rank of $M_\ell$ is $|S|-1$, then we can efficiently compute a factorization $L = L_1L_2 \ldots L_i\ldots L_c$ into pairwise relatively prime factors such that modulo each factor $L_i$, through a sequence of row operations and row/column permutations $M$ can be efficiently written (with entries modulo $L_i$) in the form 
\[ \left( \begin{array}{cccccc}
r_{L_i}(1) & * & * &   \ldots  & * & *\\
0 & r_{L_i}(2) & * &\ldots   & * &  * \\
0 &  0 & r_{L_i}(3)  &\ldots  & *  &  * \\
\vdots &  \vdots & \vdots  &\ddots  & \vdots &  \vdots \\
0 & 0 &   0  &   \ldots & r_{L_i}(|S|-1) & *\\
0 & 0 & 0 &  \ldots  & 0 & x_{L_i}(|S|)\\
\vdots & \vdots & \vdots &  \ddots  & \vdots & \vdots\\
0 & 0 & 0 &  \ldots  & 0 & x_{L_i}(|R|) \end{array} \right)\]\\
where $\forall j \in \{1,2,\ldots,|S|-1\}$, $r_{L_i}(j)$ is invertible modulo $L_i$ . 
Denote by $r_{i,j}$ the entry in the $i^{th}$ row and the $j^{th}$ column of $R$. There exists an entry in $R$ such that the set of primes dividing it is strictly contained in the set of primes dividing $L$ for otherwise there exists an $\ell$ dividing $L$ such that the $\Z/\ell\Z$ rank of $M_\ell$ is less than $|S|-1$. We may assume that this entry is $r_{1,1}$ for otherwise we may permute the rows and columns appropriately.\\ \\
If $r_{1,1}$ is not invertible modulo $L$, then we have found $\gcd(r_{1,1},L)$, a non trivial factor of $L$.  We may extract the largest factor $\hat{L}$ of $L$ supported by the primes dividing $\gcd(r_{1,1},L)$ as follows. Set $N_1 := \gcd(r_{1,1},L)$, $N_2:=\gcd(r_{1,1},L/N_1)$, $N_3:=\gcd(r_{1,1},L/(N_1N_2))$ and so on until $N_i = 1$. Then $\hat{L} = L / (N_1N_2 \ldots N_i)$. We recursively compute the desired matrix decomposition modulo $\hat{L}$ and modulo $L/\hat{L}$.\\ \\
If $r_{1,1}$ is invertible modulo $L$, then we may use it as a pivot and through row operations make every other entry in the first row zero and the resulting submatrix with the first row and column removed is of $\Z/\ell\Z$ rank $|S|-2$ for every $\ell$ dividing $L$ and is dealt with recursively.\\ \\
Since at each step, we either reduce the number of columns by $1$ or reduce into two subproblems each with modulus at most half of $L$, the number of recursive steps in our algorithm is bounded by a polynomial in $\log(L)$ and $|S|$.\\ \\
Consider when we have reduced the number of columns to $1$ (say modulo a factor $L_i$ of $L$) by performing a sequence of row operations and row/column permutations.  We have a system of relations in $(\Z S/\Z R)/L_i(\Z S/\Z R)$ of the form 
\[ \left( \begin{array}{cccccc}
r_{L_i}(1) & * & * &   \ldots  & * & *\\
0 & r_{L_i}(2) & * &\ldots   & * &  * \\
0 &  0 & r_{L_i}(3)  &\ldots  & *  &  * \\
\vdots &  \vdots & \vdots  &\ddots  & \vdots &  \vdots \\
0 & 0 &   0  &   \ldots & r_{L_i}(|S|-1) & *\\
0 & 0 & 0 &  \ldots  & 0 & x_{L_i}(|S|)  \\
\vdots & \vdots & \vdots &  \ddots  & \vdots & \vdots\\
0 & 0 & 0 &  \ldots  & 0 & x_{L_i}(N) \end{array} \right)     
\left( \begin{array}{c}
\alpha(L_i)_1\\
\alpha(L_i)_2\\
\alpha(L_i)_3\\
\vdots\\
\alpha(L_i)_{|S|-1}\\
\alpha(L_i)_{|S|}\\
\end{array} \right) =
\left( \begin{array}{c}
0\\
0\\
0\\
\vdots\\
0\\
0\\
\end{array} \right)\]\\
where $\forall j \in \{1,2,\ldots,|S|-1\}$, $r_{L_i}(j)$ is invertible modulo $L_i$ and $\alpha(L_i)_j \in (\Z S/\Z R)/L_i(\Z S/\Z R)$.\\ \\
Since $\forall j \in \{1,2,\ldots,|S|-1\}$, $r_{L_i}(j)$ is invertible modulo $L_i$, $\alpha(L_i)_{|S|}$ generates $(\Z S/\Z R)/L_i(\Z S/\Z R)$ and for all $s \in S$ we can express the representative of $s$ in $(\Z S/\Z R)/L_i(\Z S/\Z R)$ as a power of $\alpha(L_i)_{|S|}$.\\ \\
For a prime $\ell$ dividing $L_i$, $\ell$ does not divide $L/L_i$ and thus the factorization $L = \prod_i Li$ that we obtain is into relatively prime factors. Thus, by the chinese remainder theorem, $$(\Z S/\Z R)/L(\Z S/\Z R) \cong \bigoplus_i (\Z S/\Z R)/L_i(\Z S/\Z R)$$ and we can compute a generator of $(\Z S/\Z R)/L(\Z S/\Z R)$ and for every $s \in S$ compute the discrete logarithm of its image in $(\Z S/\Z R)/L(\Z S/\Z R)$ with respect to the found generator.  $\square$.
\begin{remark}\label{rem}
In lemma \ref{gaussian}, we neither assume that the factorization of $L$ is known nor assume that $S_g$ contains an element that generates $\F_g^\times[L]$. If $S_g$ does not contain an element that generates $\F_g^\times[L]$, then $M$ does not have a $|S|-1$ by $|S|-1$ submatrix whose $\Z/\ell\Z$ rank is $|S|-1$ for all prime $\ell \mid L$. Thus, if $S_g$ does not contain an element that generates $\F_g^\times[L]$, the factorization $L =\prod_i L_i$ that results from the proof of the lemma is non-trivial.
\end{remark}
\begin{theorem}\label{theorem_g} Let $L$ be a positive integer such that $L \mid |\F_g^\times|$ and $\gcd(L,|\F_{g}^\times|/L)=1$. If for all prime $\ell$ dividing $L$, the $\Z/\ell\Z$ rank of $\Z R \otimes \Z/\ell\Z$ is $|S|-1$, then in time polynomial in $\log(L)$, $|R|$ and $|S|$,  we can find a generator $\beta_L$ of $\F_g^\times[L]$ and compute the discrete logarithm of the projection in $\F_g^\times[L]$ of every element in $S_g$with respect to $\beta_L$. 
\end{theorem}
\ \\ For a prime $\ell$, the $\Z/\ell\Z$ rank of $\Z R \otimes \Z/\ell\Z$ is the same as the $\Z/\ell\Z$ rank of the matrix $M_\ell$. If for all prime $\ell$ dividing $L$, the $\Z/\ell\Z$ rank of $M_\ell$ is $|S|-1$, then by lemma \ref{gaussian} in time polynomial in $\log(L)$, $|R|$ and $|S|$, we can find a generator of $(\Z S/\Z R)/L(\Z S/\Z R)$ and for every $s \in S$ compute the discrete logarithm of its image in $(\Z S/\Z R)/L(\Z S/\Z R)$ with respect to the found generator. Hence we can compute under the surjection $$ (\Z S/\Z R)/L(\Z S/\Z R) \twoheadrightarrow \F_g^\times/(\F_g^\times)^L$$ a generator and with respect to it the discrete logarithms of the image in $\F_g^\times/(\F_g^\times)^L$ of every element in $S_g$.\\ \\
Since $\gcd(L,|\F_{g}^\times|/L)=1$, $\F_g^\times[L] \cong \F_g^\times/(\F_g^\times)^L$ and the theorem follows. $\square$.\\ \\
In \S~\ref{dlog_snf}, we present an alternate proof of theorem \ref{theorem_g} using Smith normal form computation over $\Z$. 
\subsection{Discrete Logarithms in the Factorbase Using Smith Normal Forms}\label{dlog_snf}
\ \\In this section, we present an alternate proof of theorem \ref{theorem_g} which we restate below for convenience.
\begin{theorem}\label{theorem_g_snf} Let $L$ be a positive integer such that $L \mid |\F_g^\times|$ and $\gcd(L,|\F_{g}^\times|/L)=1$. If for all prime $\ell$ dividing $L$, the $\Z/\ell\Z$ rank of the $\Z/\ell\Z$ module generated by $R_\ell$ is $|S|-1$, then in time polynomial in $\log(L)$, $|R|$ and $|S|$,  we can find a generator of $\beta_L$ for $\F_g^\times[L]$ and compute the discrete logarithm of the projection in $\F_g^\times[L]$ of every element in $S_g$with respect to $\beta_L$. 
\end{theorem}
\ \\Since $R$ is a set of relations for $\langle S_g \rangle$, we have the natural surjection $$\varphi : \Z S/\Z R \twoheadrightarrow \langle S_g \rangle$$
$$\ \ \ \ \ \  \ \ \ \ \ \ \ \ \ \ \ \ \ \ \ \ \ \sum_{s \in S}z_s s + \Z R  \longmapsto \prod_{s \in S} s(x)^{z_s} \mod g(x).$$
The Smith normal form of of the relation matrix $M$ gives the decomposition of $\Z S/\Z R$ into invariant factors $$\Z S/\Z R = \langle e(1) \rangle \oplus \langle e(2) \rangle \oplus \ldots \oplus \langle e(|S|) \rangle  \cong  \Z/d_1\Z \oplus \Z/d_2\Z \oplus \ldots \oplus \Z/d_{|S|}\Z $$
where for $1\leq i \leq |S|$, $e(i) \in \Z S/\Z R$ and $d_i$ denotes the order of $e(i)$ in $\Z S/\Z R$ and for $1 \leq i <|S|$, $d_i \mid d_{i+1}$.\\ \\
The condition that for all prime $\ell \mid L$, the $\Z/\ell\Z$ rank of the $\Z/\ell\Z$ module generated by $R_\ell$ is $|S|-1$ implies that for all prime $\ell \mid L$, $(\Z S/\Z R)/\ell(\Z S/\Z R)$ is cyclic and non-trivial. Hence, for all prime $\ell \mid L$, $(\Z S/\Z R)(\ell)$ is cyclic. Further, under the surjection $(\Z S/\Z R)(\ell) \twoheadrightarrow \langle S_g \rangle(\ell)$, a generator of $(\Z S/\Z R)$ maps to a generator of $\F_g^\times(\ell)$.
Thus for all prime $\ell$ dividing $L$,  the projection of $S_g$ in $\F_g^\times(\ell)$ generates $\F_g^\times(\ell)$. We may thus conclude that $L$ divides $d_{|S|}$.\\ \\
Since $L$ divides the order of $\varphi(e(|S|))$, the projection (call $\beta_L$) of $\varphi(e(|S|))$ in $\F_g^\times[L]$ generates $\F_g^\times[L]$.\\ \\
Let $ \vartheta : \Z S/\Z R \twoheadrightarrow \langle e(|S|) \rangle $ denote the natural surjection from $\Z S/\Z R$ to its largest invariant factor.
Let $\pi$ denote a projection from $\F_g^\times$ to $\F_g^\times[L]$. Then we have $$\Z S/\Z R \xrightarrow{\ \ \vartheta  \ \ } \langle e(|S|)\rangle \xrightarrow{\ \ \ \  \pi \circ \varphi\ \ \ \  } \F_{g}^\times[L]$$
$$\ \ \ \ \ \ \ \ \ \ \ \ \ \ \ \ \ \ \kappa \longmapsto e(|S|)^{\theta(\kappa)} \longmapsto \left(\bar{\varphi}(e(|S|))\right)^{\theta(\kappa)}.$$
Given a $\Z S$ representative of an element $\kappa \in \Z S/\Z R$, the Smith normal form of $M$ allows us to efficiently compute an integer $\theta(\kappa)$ such that $\vartheta(\kappa) = e(|S|)^{\theta(\kappa)}$.\\ \\
Let the images of $\sum_{s \in S}a_s s$ and $\sum_{s \in S}c_s s$ in $\Z S/\Z R$ be $\kappa_1$ and $\kappa_2$ respectively. Then $\pi(\prod_{s \in S} s(x)^{a_s}\mod g(x)) = \left(\bar{\varphi}(e(|S|))\right)^{\theta(\kappa_1)}$ and $\pi(\prod_{s \in S} s(x)^{c_s}\mod g(x)) =\left(\bar{\varphi}(e(|S|))\right)^{\theta(\kappa_2)}$.\\ \\
Since $L$ divides the order of $\varphi(e(|S|))$, $\langle \bar{\varphi}(e(|S|))\rangle = \F_g^\times[L]$.\\ \\
Thus, $\pi(\prod_{s \in S} s(x)^{a_s} \mod g(x))$ is in the subgroup generated by $\pi(\prod_{s \in S} s(x)^{c_s} \mod g(x))$ if and only if there exists an integer $j$ such that $$\theta(\kappa_1) = j \theta(\kappa_2) \mod L.$$
If such a $j$ exists, then $j \mod L$ is the discrete logarithm of the projection of $\prod_{s \in S} s(x)^{a_s} \mod g(x)$ in $\F_g^\times[L]$ with respect to the projection of $\prod_{s \in S} s(x)^{c_s} \mod g(x)$ in $\F_g^\times[L]$ as the base.\\ \\
We can decide if such a $j$ exists and if so find one using the extended Euclidean algorithm. $\square$.

\subsection{Algorithms for the Descent}\label{algorithm_descent}

\ \\ \ \\ Let $U \subset \F[x]$ and $V \subset \F[x]$ be finite disjoint sets and let $U_h$ and $V_h$ respectively denote the set of elements of $U$  and  of $V$ modulo $h(x)$. Consider an algorithm that tries to descend from $V$ to $U$ by generating a set $T \subset \Z (V\cup U)$ of relations where every $(\sum_{v \in V} a_v + \sum_{u \in U} a_u) \in T$ satisfies $$\prod_{v \in V} v(x)^{a_v} = \prod_{u \in U} u(x)^{a_u}  \mod h(x).$$
Recall that for $r=\sum_{v \in V}b_v v + \sum_{u \in U} b_u u \in T$,  $r_V$ denotes $\sum_{v \in V}b_v v$ and $T_V$ denotes $\{r_V| r \in T\}$. Further, from the definition, $V$ is descent to $U$ through $T$ if and only if $\Z T_V = \Z V$. Likewise for a positive integer $L$, $V$ to be descent to $U$ through $T$ modulo $L$ if and only if $\Z T_V  \otimes \Z/L\Z= \Z V \otimes \Z/L\Z$. Hence we have the following lemma.

\begin{lemma}\label{descent_necessary} For a prime $\ell$, descent from $V$ to $U$ modulo $\ell$ through $T$ is possible if and only if the $\Z/\ell\Z$ rank of $\Z T_V  \otimes \Z/\ell\Z$ is $|V|$ if and only if $(\Z S/\Z T_V) \otimes Z/\ell\Z$ is trivial.
\end{lemma}
\ \\With straightforward modifications, the algorithms developed to prove lemma \ref{gaussian} and theorem \ref{theorem_g} apply to the descent phase as well and lead to theorem \ref{theorem_gg}. Let $\bar{U}$ denote the subset of $U$ where an element is in $\bar{U}$ if and only if it appears with a non zero coefficient in at least one relation in $T$.
\begin{theorem}\label{theorem_gg} Let $L$ be a positive integer such that $L \mid |\F_g^\times|$ and $\gcd(L,|\F_g^\times|/L)=1$. If for all prime $\ell$ dividing $L$, the $\Z/\ell\Z$ rank of $\Z T_V \otimes \Z/\ell\Z$ is $|V|$, then given $T$, in time polynomial in $\log(L),|\bar{U}|,|V|$ and $|T|$, we can descend from $V$ to $U$ in $\F_g^\times[L]$. That is, for every $v \in V$, we can efficiently find $w \in \Z$ and $(w_b)_{b \in U} \in \Z U$ such that the projections of $v(x)^{w}$ and $\prod_{b \in U} b(x)^{w_b} $ to $\F_h^\times[L]$ are identical.
\end{theorem}

\subsection{Implications on Polynomial Selection}\label{poly_select}
\ \\In this section, we impose conditions on $h(x)$ and show that under these conditions, relations for $\F_h^\times$ suffice in efficiently computing discrete logarithms in $\F_g^\times$.\\ \\
Fix a positive integer $C$ that defines a smoothness bound. We say that an integer is $|\F|^{C}$-smooth if and only if all its prime factors are at most $|\F|^{C}$. For this subsection, let $v$ denote the largest factor of $|\F_g^\times|$ that is $|\F|^{C}$-smooth and let $L := |\F_g^\times|/v$. Both $v$ and $L$ can be efficiently computed from the knowledge of $|\F|,C$ and $\deg(g)$. Since $L$ and $v$ are relatively prime, $$\F_g^\times = \F_g^\times[v] \times \F_g^\times[L]$$
and we can project from $\F_g^\times$ to $\F_g^\times[v]$ by taking $L^{th}$ powers. Since the order of $\F_g^\times[v]$ is $|F|^{C}$-smooth, the discrete logarithm problem in $\F_g^\times[v]$ can be solved in time polynomial in ${|\F|^C}$ using the Pohlig-Hellman algorithm \cite{ph}. All that remains is to address the discrete logarithm computation in $\F_g^\times[L]$.\\ \\ 
We first insist that $(g(x))^2$ does not divide $h(x)$. Let $h(x) = g(x) \prod_{i=1}^k g_i(x)^{a_i}$ be a factorization where $g_i(x)$ are distinct irreducible polynomials in $\F[x]$. The chinese remainder theorem over $\F[x]$ implies $$\F_h^\times= \prod_{i=0}^k \F_{g_i^{a_i}}^\times.$$
The orders of the groups in $\{\F_{g_i^{a_i}}^\times\ |\ 0 \leq i \leq k\}$ are not relatively prime since every $\F_{g_i^{a_i}}^\times$ contains $\F^\times$ as a subgroup. Thus $\F_h^\times$ is not cyclic and by theorem \ref{obstruction_theorem} there is an obstruction to either computing discrete logarithms in the factorbase or for the descent.\\ \\
However since $|\F|$ is $|\F|^C$-smooth, the discrete logarithm computation in $\F_h^\times[|\F|]$ is dealt with by the Pohlig-Hellman algorithm.\\ \\
The concern is when $h(x)$ has a factor $g_i(x)$ other than $g(x)$ such that $|\F_g^\times|$ and $|\F_{g_i}^\times|$ share a large prime factor $\ell$. In this case, by theorem \ref{obstruction_theorem} there is an obstruction modulo a large prime $\ell$ to either computing discrete logarithms in the factorbase or for the descent.\\ \\
For instance when there is a $j$ such that the degrees of $g_j(x)$ and $g(x)$ share a large enough factor, the existence of a large prime factor dividing both $|\F_{g}^\times|$ and $|\F_{g_j^{a_j}}^\times|$ is all but certain.\\ \\ 
To avoid this, we impose a second condition on $h(x)$ and insist that $\gcd(\left|\F_{h/g}^\times \right|, |\F_g^\times|)$ is $|\F|^{C}$-smooth.\\ \\
By our choice of $h(x)$, for a non $|\F|^C$-smooth prime $\ell$ dividing $\F_g^\times$, $\ell$ does not divide $|\F_{h/g}^\times|$ and it follows that $\F_h^\times/(\F_h^\times)^\ell \cong \F_g^\times/(\F_g^\times)^\ell$ and thus $\F_h^\times/(\F_h^\times)^\ell$ is cyclic since $\F_g^\times/(\F_g^\times)^\ell$ is cyclic. Hence $\F_h^{\times}(\ell)$ is cyclic and the obstructions do not arise.

\section{Applications to Jouxs Relation Generation Algorithm.}\label{obstruction_joux}
\ \\The results derived in \S~\ref{residue_class_rings} can be applied to the algorithms in \cite{jl,gggz,joux,bgjt} in solving discrete logarithms in a finite extension of the finite field $\F$. We illustrate this in this section by describing the application to the algorithm of Joux \cite{joux} for computing discrete logarithms in the factorbase.  Applications to the descent of Barbulescu-Gaudry-Joux-Thome \cite{bgjt} are described in \S~\ref{obstruction_descent}.\\ \\ 
Joux's algorithm proceeds by embedding $\F_{p^n}$ into an extension $\F_{q^{2n}}$ where $q$ is a power of $p$ such that $n \leq q$. 
Polynomials $h_0(x), h_1(x) \in \F_{q^2}[x]$ of low degree such that the factorization of $ h_1(x)x^q - h_0(x)$ over $\F_{q^2}[x]$ has an irreducible factor of degree $n$ are then sought. If found, one such irreducible factor of degree $n$ is picked as $g_0(x)$. The field $\F_{q^{2n}}$ is constructed as $\F_{q^2}[x]/g_0(x)\F_{q^2}[x]$. Due to Lenstra \cite{lenstra}, an isomorphism between two explicit representations of a finite field can be efficiently computed. As a consequence, the fact that we work over a specially chosen representation of the finite field which may differ from the input representation wherein the discrete logarithm is to be solved is not a concern. The motivation behind choosing $g_0(x)$ in this manner is that the identity $h_1(x) x^q - h_0(x)=0 \mod g_0(x)$ is used by the relation generation algorithm to replace $x^q \mod g_0(x)$ with an expression consisting of the low degree polynomials $h_0(x)$ and $h_1(x)$ modulo $g_0(x)$.\\ \\ 
To apply results of \S~\ref{residue_class_rings} in this context, we set $\F=\F_{q^2}, g(x)=g_0(x),h(x) = h_1(x)x^q-h_0(x)$ and take $S$ to be the set of monic linear polynomials in $\F_{q^2}[x]$ along with $h(x)$ and a generator $\lambda$ of $\F_{q^2}^\times$. Further, \cite[Thm. 8]{chung}\cite[Ques 1.1]{wan} guarantees that $\langle S_g \rangle \cong \F_g^\times \cong \F_{q^{2m}}^\times$. The relation generation phase collects relations for $\langle S_h \rangle$.\\ \\
If $h(x)/g(x)$ were to have a factor $\bar{g}(x)$ other than $g(x)$ such that $|\F_g^\times|$ and $|\F_{\bar{g}}^\times|$ share a large prime factor $\ell$, then $\F_h^\times(\ell)$ is not cyclic and thus there is an obstruction either to determining discrete logarithms in the factorbase or to the descent to the factorbase. We next argue that the former is likely to occur. Since $\deg(\bar{g}) \leq q$, \cite[Thm. 8]{chung}\cite[Ques 1.1]{wan} implies that $\langle F_{\bar{g}} \rangle \cong \F_{\bar{g}}^\times$. Thus $\langle F_h \rangle(\ell)$ projects to both $\F_g^\times(\ell)$ and $\F_{\bar{g}}^\times(\ell)$ surjectively. Unless the set of relations for $\F_g^\times(\ell)$ and $\F_{\bar{g}}^\times(\ell)$ are identical (which is unlikely), the $\ell$-primary part of $\langle F_h\rangle$ is not cyclic.\\ \\
We note that in the context of Joux's algorithm, the obstructions described in \S~\ref{residue_class_rings} are easy to resolve in the Kummer case, that is when $n=q-1$. When $n=q-1$, $h(x)$ is chosen as $x^q-\lambda x$ and $g(x)$ as $x^{q-1}-\lambda$ where $\lambda$ is a generator of $\F_{q^2}^\times$. In this case, $h(x) = xg(x)$ and thus $(\F_h^\times)^{q^2-1}$ is cyclic. Further, the relation $x^{q-1} = \lambda  \mod g(x)$ can be added to the relation matrix and this allows the inclusion of $x \mod g(x)$ in the factorbase.\\ \\
We propose the following modification to the polynomial selection phase of Joux's algorithm to avoid the obstruction. We embed $\F_{p^n}$ into an extension $\F_{q^{2m}}$ where $q$ is a power of $p$ such that $n \leq q$ and $m$ is a multiple of $n$ such that $q/2 < m \leq q$. In particular, we set $q := p^{\lceil \log_p(n)\rceil}$ and $m$ is chosen as the largest integral multiple of $n$ satisfying $q/2 < m \leq q$. We  fix $C$ to be a positive integer constant independent of $p$ and $n$ and insist that $h(x)$ in addition to having a proper irreducible factor $g(x)$ of degree $m$, satisfies the following three conditions. 
\begin{enumerate}
\item The square of $g(x)$ does not divide $h(x)$.
\item $\gcd(\left|\F_{h/g}^\times \right|, |\F_g^\times|)$ is $|\F|^{C}$-smooth.
\item $h(x)$ does not have linear factors.
\end{enumerate}
If $h(x)$ were to have a linear factor, then the relation generation step will not relate that linear factor to the rest of the linear polynomials in the factor base since the image of the linear factor is not in the unit group $\F_h^\times$. As a result, we would have to exclude that linear factor from the factor base and F.R.K Chung's theorem that ensures  $\langle S_g \rangle \cong \F_g^\times$ would no longer apply. It is to circumvent this that we insisted that $h(x)$ have no linear factors.\\ \\
We call a choice of $h(x)$ that satisfies the above conditions as $C$-$good$. A formal definition of $C$-good is in \S~\ref{poly_search}. The modified polynomial search algorithm follows.\\ \\
\ \\ \textbf{Search for $h_0(x), h_1(x)$ and $g(x)$: }\textit{Fix positive integers $C,D$. Enumerate candidates for $h_0(x), h_1(x) \in \F_{q^2}[x]$ with each of their degrees bounded by $D$. For each candidate pair $(h_0(x) , h_1(x))$, factor $h(x) = h_1(x)x^q-h_0(x)$. If $h(x)$ is $C$-$good$, output $h_0(x), h_1(x)$ and the factor of degree $m$ and stop. If no such candidates are found, declare failure.}\\ \\
The search algorithm terminates after considering at most $q^{2(D+1)} = q^{\mathcal{O}(1)}$ candidate pairs. Factoring each candidate $h_1(x)x^q-h_0(x)$ takes time polynomial in the degree $q+D$ and $p$ using Berlekamp's deterministic polynomial factorization algorithm \cite{ber}. Given the degrees of the irreducible factors and the corresponding powers in the factorization of $h(x)$, we can efficiently test if it is $C$-$good$. Thus, the search for $h_0(x), h_1(x)$ and hence $g(x)$ of the desired form takes at most $q^{\mathcal{O}(1)}$ time.\\ \\
Our choice of embedding field $\F_{q^{2m}}$ is in certain cases larger than the field $\F_{q^{2n}}$ chosen in Joux's algorithm \cite{joux}. 
In \S~\ref{poly_search}, under assumptions similar to those made in Joux's polynomial search, we arrive at the following heuristic which asserts that it is sufficient for $C$ and $D$ to be constants independent of $q$ and $n$ to ensure the success of the the modified polynomial search.
\begin{heuristic}\label{heu1} There exists positive integers $C,D$ such that for all prime powers $q$ and for all positive integers $2 < m \leq q$, there exists $h_0(x),h_1(x) \in  \F_{q^2}[x]$ of degree bounded by $D$ such that $h_1(x)x^q-h_0(x)$ is $C$-$good$.
\end{heuristic}
\ \\ For the success of later steps in Joux's relation generation\cite{joux}, it is critical that the degree bound $D$ is a constant independent of $q$ and $m$. Our modified polynomial search does ensure that $D=\mathcal{O}(1)$.\\ \\
We next turn to computing discrete logarithms from the relations obtained from Joux's algorithm with the modified polynomial selection.
For the rest of this section, let $L$ denote the order of the non-smooth component of $\F_g^\times$. That is $L$ is $|\F_g^\times|$ divided by the largest $|\F|^C$-smooth factor of $|\F_g^\times|$. Following the discussion in \S~\ref{poly_select}, we only need to address the discrete logarithm computation in $\F_g^\times[L]$.\\ \\
Since $\gcd(\left|\F_{h/g}^\times \right|, |\F_g^\times|)$ is $|F|^{C}$-smooth for our chosen $h(x)$, for a prime $\ell$ dividing $L$, $\F_h^{\times}(\ell)$ is cyclic and hence obstructions do not arise for the non-smooth component thereby suggesting the following heuristic.
\begin{heuristic}\label{heuristic_factorbase}
Let $R \subset \Z S$ denote the set of relations collected by Joux's relation generation algorithm with the polynomial $h(x)$ chosen in accordance with the modified polynomial selection. Let $L$ denote $|\F_g^\times|$ divided by the largest $|\F|^C$-smooth factor of $|\F_g^\times|$. Then, for every prime $\ell$ dividing $L$, the $\Z/\ell\Z$ rank of $\Z R \otimes \Z/\ell\Z$ is $|S|-1$. 
\end{heuristic}
\ \\ The discussion in \S~\ref{poly_select}, the Pohlig-Hellman algorithm and theorem \ref{theorem_g} together imply the following theorem.
\begin{theorem}\label{dlog_factorbase_theorem}
Under heuristics \ref{heu1} and \ref{heuristic_factorbase}, there is a deterministic algorithm that in time polynomial in $q$ and $m$ finds a generator of $\F_{g}^\times \cong \F_{q^{2m}}^\times$ and with respect to it computes the discrete logarithm of every element in $S_g$.
\end{theorem}

\subsection{Polynomial Search}\label{poly_search}
For a positive integer $C$, we formally define a polynomial $f(x) \in \F_{q^2}[x]$ to be $C$-$good$ if and only if the following four conditions are satisfied.
\begin{enumerate}
\item $f(x)$ has an irreducible factor of degree $m$ (call it $g(x)$).
\item The square of $g(x)$ does not divide $f(x)$.
\item $f(x)$ does not have linear factors.
\item $\gcd(\left|\F_{f/g}^\times \right|, |\F_g^\times|)$ is $|\F|^C$-smooth.
\end{enumerate}
We set a degree bound $D$ and investigate the existence of $h_0(x),h_1(x) \in \F_{q^{2}}[x]$ each of degree bounded by $D$ such that $h(x) = h_1(x) x^q-h_0(x)$ is $C$-$good$. \\ \\
The existence of $C$-$good$ polynomials of the above form requires that $q+D$ is at least $m+2$  for otherwise we are left with a linear factor. To this end, if $m=q$, we assume $D>1$ and if $m=q-1$, we assume $D>0$.\\ \\
We next state for each condition, a probability estimate that a random polynomial of degree $q+D$ satisfies it.\\ \\
The fraction of polynomials in $\F_{q^2}[x]$ of degree $q+D$ that are square free is $1-1/q^2$ \cite{panario}. Thus condition $2$ is satisfied by a random polynomial of degree $q+D$ with probability close to $1$. 
The probability that a random polynomial of degree $q+D$ does not have linear factors is $q^{2(q+D)}(1-1/q^2)^{q^2}$ which for large $q$ tends to $1/e \approx 0.36$ \cite{panario}.
The probability that a random polynomial of degree $q+D$ has an irreducible factor of degree $m$ is $1/m$ \cite{panario}.
It is thus heuristically likely that a random polynomial of degree $q+D$ satsifies the first three conditions is $\Theta(1/m)$.\\ \\
The average number of factors of a degree $q+D$ polynomial is $\mathcal{O}(\log(q+D))$ with a standard deviation bounded by $\mathcal{O}(\sqrt{\log(q+D)})$ \cite{panario}. For a polynomial that satisfies the first three conditions, it is thus heuristically likely that each of its factors excluding its degree $m$ factor is either of degree $m^\prime$ which is prime to $m$ or bounded by $C$. If that is the case, then for every factor of degree $m^\prime \neq m$, heuristically $q^{2m}-1$ and $q^{2m^\prime-1}$ are likely to be $|\F|^C$-smooth. \\ \\
If we were to assume that a random polynomial of the form $h_1(x)x^q - h_0(x)$, where $h_0(x)$ and $h_1(x)$ are of degree at most $D$ behaves like a random polynomial of the same degree, then it is $C$-$good$ with probability $\Omega(1/m)$. We can conclude heuristically that choosing $D = \Theta(\log_{q^2}(m)) = \mathcal{O}(1)$ and $C =\Theta(1)$ are sufficient to guarantee that the search succeeds in finding a $C$-$good$ polynomial that we seek and heuristic \ref{heu1} follows.

\section{Applications to the Barbulescu-Gaudry-Joux-Thome Descent.}\label{obstruction_descent}


\ \\The polynomial $h(x)$ and the factorbase $S$ chosen in the \cite{bgjt} descent are identical to the choices made in Joux's algorithm. The descent step takes an element in $\F[x]$ of degree at most $\deg(g)$ as input and attempts to express it as a product of powers of elements in the factorbase $S$ modulo $g(x)$. Thus if the descent succeeds, then we would have solved the discrete logarithm problem in $\F_g^\times$ assuming that the discrete logarithms in $S_g$ are already determined by Joux's algorithm.\\ \\
To apply results of \S~\ref{residue_class_rings} to the descent, in this section, we set $\F=\F_{q^2}$ and take $g(x)$, $h(x)$ and $S$ as chosen by Joux's algorithm with the modified search phase described in \S~\ref{obstruction_joux}.\\ \\
We first outline the descent algorithm starting from a $P(x) \in \F[x]$ of degree $w$ where $1<w<\deg(g)$. The first step attempts to reduce the problem to performing a descent on a set of inputs each of degree $w/2$ or less. To this end, a set of relations modulo $h(x)$ relating the set $V_w:=\{ P(x)+\alpha|\alpha \in \F \}$ of $\F$ translates of $P(x)$ with the set $U_w$ polynomials of degree at most $w/2$ are obtained. Let $A_w$ denote the polynomials in $U_w$ that appear in the relations obtained. From the relations obtained, we then attempt to express modulo $h(x)$ each element in $V_w$ as a product of powers of elements in $A_w$.\footnote{$U_w$ also contains $h_1(x)$ and a generator of $\F^\times$, a fact we ignore for ease of exposition.} \\ \\
The algorithm then recursively performs the descent step on every polynomial in $A_w$, that is on the polynomials of degree at most $w/2$ that appear in the first step.

\ \\ In \cite{cwz}, the following scenario was identified as a possible trap that prevents a descent step from working. Consider a step in the descent where the following multiplicative identity in $\F_h$
\begin{equation}\label{trap_informal} \prod_{v \in V_w} (v(x))^{e_v}  =  \prod_{u \in A_w} u(x)^{e_u} \mod h(x) \end{equation} 
is collected, where a $u(x) \in U_w$ appears such that $u(x)$ shares a non trivial factor with $h(x)$ and $e_u \neq 0$. For such a $u(x)$ to appear, it is necessary that there is an element in $V_w$ that is not a unit in $\F_h^\times$. In the next step, one tries to relate $u(x)$ and its $\F$ translates modulo $h(x)$ to powers of irreducible polynomials of degree at most half of $\deg(v)$. However, since $u(x)$ is irreducible in $\F[x]$ and not a unit modulo $h(x)$, $u(x)$ would never appear in a relation in $\F_h^\times$ involving only the $\F$ translates of $u(x)$ and smaller degree polynomials.\\ \\ 
To remedy this scenario, it was proposed in \cite{cwz} to not use relations where in such a $u(x)$ is involved. As a result the necessity to perform a descent on $u(x)$ would not arise. This trap avoidance strategy comes at a cost since certain relations are not utilized.\\ \\
We propose an alternative way to deal with the traps which we next allude to. While it is true due to zero divisors that the multiplicative identity \ref{trap_informal} in $\F_h$ does not yield a relation in the unit group $\F_h^\times$, it does yield a relation in $\F_g^\times$ which is the unit group we are ultimately interested in. For a descent step, let $\hat{h}(x) \in \F[x]$ denote the largest factor of $h(x)$ such that every collected multiplicative identity in $\F_h$ yields a relation for $\F_{\hat{h}}$. Note that $g(x) \mid \hat{h}(x) \mid h(x)$. Further, since the number of factors of $h(x)$ is small compared to the number of relations we expect to get, in addition to $V_w$, we can try to eliminate the factors of $h(x)$ that appear in the collected multiplicative identities in $h(x)$. Thereby in the recursive steps that follows, we never have to descend starting from a zero divisor in $\F_h$.\\ \\
The advantage to our approach of handling traps is that we break the symmetry between $g(x)$ and the factors of $h(x)/\hat{h}(x)$ by finding relations that hold in $\F_{\hat{h}}^\times$. Thereby in a descent step, instead of trying to descend through relations for $\F_h^\times$, we descend through relations for $\F_{\hat{h}}^\times$. Unless $\hat{h}(x)$ happens to be $g(x)$, we still resort to the results in \S~\ref{residue_class_rings} to efficiently perform the descent.\\ \\
A formal description of the \cite{bgjt} descent and our proposed modification follows.\\ \\
An element $\eta \in \F_g^\times$ is presented to the descent algorithm as a polynomial $P(x) \in \F_{q^2}[x]$ of degree $w$ such that $\eta = P(\zeta)$ and $w <m$.\\ \\
We may assume that $P(x)$ and $h(x)$ do not share a non constant factor. Otherwise, raise $P(x)$ to a random power, then divide by $h(x)$ and call the remainder $P^\prime(x)$. It is likely that $P^\prime(x)$ and $h(x)$ do not share a factor and hence we can start the descent from $P^\prime(x)$.\\ \\
The first step in the descent attempts to reduce the problem to performing a descent on a set of inputs each of degree $w/2$ or less. To this end, a set of multiplicative relations modulo $h(x)$ relating the $\F_{q^2}$ translates of $P(x)$ with polynomials of degree at most $w/2$ are obtained. From the relations obtained, we then attempt to express modulo $h(x)$ each $\F_{q^2}$ translate of $P(x)$ as a product of powers of polynomials of degree at most $w/2$ and powers of $\lambda$ and $h_1(x)$.\\ \\
The first step starts with the identity $$\prod_{\alpha \in \F_{q}}{x-\alpha} = x^q - x.$$
Let $\mathcal{P}_q \subset GL(2,q)$ be a set of representatives of the left cosets of $PGL(2,q)$ in $PGL(2,q^2)$.\\ \\
For \[\mathfrak{m} = \left( \begin{array}{cc}
a & b  \\
c & d   \end{array} \right) \in \mathcal{P}_q, \]
the substitution $x \mapsto \frac{aP(x)+b}{cP(x)+d}$ yields\\ $$ \prod_{\alpha \in \F_q} \frac{(a-\alpha c)P(x) + (b-\alpha d)}{(c P(x) +d)^{q}} = \frac{(c P(x) + d)(a P(x) + b)^q - (aP(x)+b)(cP(x)+d)^q }{(cP(x)+d)^{q+1}}$$  
$$\Rightarrow (cP(x)+d) \prod_{\alpha \in \F_q} ((a-\alpha c)P(x) + (b-\alpha d)) = (c P(x) + d)(a P(x) + b)^q - (aP(x)+b)(cP(x)+d)^q. $$
Linearity of raising to the $q^{th}$ power implies $$ (cP(x)+d) \prod_{\alpha \in \F_q} ((a-\alpha c)P(x) + (b-\alpha d)) = (c P(x) + d)(a^q \tilde{P}(x^q) + b^q) - (a P(x)+b)(c^q \tilde{P}(x^q)+d^q).$$ where $\tilde{P}(x)$ is $P(x)$ with its coefficients raised to the $q^{th}$ power.\\ \\
By substituting $x^q = \frac{h_0(x)}{h_1(x)}$, we obtain a congruence module $h(x)$. Under the substitution, the right hand side becomes $$(c P(x) + d)(a^q \tilde{P}\left(\frac{h_0(x)}{h_1(x)}\right) + b^q) - (a P(x)+b)(c^q \tilde{P}\left(\frac{h_0(x)}{h_1(x)}\right)+d^q) $$ which can be expressed as a fraction $$N_{\mathfrak{m},P}(x)/D_{\mathfrak{m},P}(x)$$ where $N_{\mathfrak{m},P}(x) \in \F_{q^2}[x]$ is of degree bounded by $(1+D)w$ and $D_{\mathfrak{m},P}(x) \in \F_{q^2}[x]$ is a power of $h_1(x).$\\ \\
If $N_{\mathfrak{m},P}(x)$ were to factor over $\F_{q^2}[x]$ into a product of irreducible factors each of degree bounded by $w/2$, then we obtain a relation of the form $$ \prod_{\beta \in \F_{q^2}} (P(x)-\beta)^{e_\beta}  =  \lambda^{b_\lambda}h_1(x)^{e_{h_1}} \prod_{u \in W_{\mathfrak{m},P}} u(x)^{b_u} \mod h(x)$$
where $\forall \beta \in \F_{q^2}, e_\beta \in \{0,1\}$ (See \cite{bgjt} for a proof), $W_{\mathfrak{m},P}$ denotes a subset of the set of monic irreducible polynomials in $\F_{q^2}[x]$ of degree bounded by $w/2$, $b_\lambda,b_{h_1} \in \Z$ and $\forall u \in W_{\mathfrak{m},P},\ b_u \in \Z-\{0\}$.\\ \\
Let $W_P$ denote the union of the sets $W_{\mathfrak{m},P}$ as $\mathfrak{m}$ ranges over elements in $\mathcal{P}_q$ that result in a relation.\\ \\
We next attempt to descend from $P(x) + \F_{q^2}$ to $W_P$ through the set of relations generated. We recursively perform the descent on the elements in $W_P$ until we decompose into linear factors.\\ \\ 
In \cite{cwz}, the following scenario was identified as a possible trap that prevents a descent step from working. Consider a $\mathfrak{m} \in \mathcal{P}_q$ that results in the following relation
\begin{equation} \prod_{\beta \in \F_{q^2}} (P(x)-\beta)^{e_\beta}  =  \lambda^{b_\lambda}h_1(x)^{e_{h_1}} \prod_{u \in W_{\mathfrak{m},P}} u(x)^{b_u} \mod h(x) \end{equation}
where in a $v(x) \in U_{\mathfrak{m},P}$ appears such that $v(x)$ divides $h(x)$. In the next step, one tries to relate $v(x)$ and its $\F_{q^2}$ translates modulo $h(x)$ to powers of irreducible polynomials of degree at most half of $\deg(v)$. However, since $v(x)$ is irreducible in $\F_{q^2}[x]$ and not a unit modulo $h(x)$, $v(x)$ would never appear in a relation in $\F_h^\times$ involving only the $\F_{q^2}$ translates of $v(x)$ and smaller degree polynomials. The trick of raising $v(x)$ to a random power modulo $h(x)$ is not available \footnote{In \cite{bgjt2}[Prop 10], a descent step starting from $v(x)$ for the case when $D \leq 2$ is described.} in the intermediate steps since it might raise the degree.\\ \\
To remedy this scenario, it was proposed in \cite{cwz} to not use relations where in such a $v(x)$ is involved. As a result the necessity to perform a descent on $v(x)$ would not arise. This trap avoidance strategy comes at a cost since certain relations are not utilized.\\ \\
The modification to the descent step we propose is that at each step we attempt to express every element in $$V_P := \left\{\frac{P(x)-\beta }{\gcd(P(x)-\beta,h(x)/g(x))} | \beta \in \F_{q^2}\right\} \bigcup G_P$$ modulo $g(x)$ as a product of powers of polynomials of degree at most $w/2$ and powers of $h_1(x)$ and $\lambda$. Here $G_P$ is the set of all factors of $h(x)/g(x)$ that appear in the descent step involving $P(x)$. A formal definition of $G_P$ is in the description below.\\ \\
Say $N_{\mathfrak{m},P}$ does factor over $\F_{q^2}[x]$ into a product of irreducible factors each of which is either of degree bounded by $w/2$ or a factor of $h(x)$.\\ \\ 
The image of every factor of $h(x)/g(x)$ in $\F_g$ is a unit and hence can be inverted resulting in a relation of the form
$$\left( \prod_{i=1}^k g_i(x)^{s_{\mathfrak{m},i}}\right) \times \prod_{\beta \in \F_{q^2}} \left( \frac{P(x)-\beta}{\gcd(P(x)-\beta,h(x)/g(x))}\right)^{r_{\mathfrak{m},\beta}}   =  \lambda^{c_{\mathfrak{m},\lambda}} h_1(x)^{c_{\mathfrak{m},h_1}} \prod_{u \in U_{\mathfrak{m},P} } u(x)^{c_{\mathfrak{m},u}} \mod g(x)$$ where $U_{\mathfrak{m},P}$ is a set of monic irreducible polynomial of degree at most $w/2$ each of whose elements is not a factor of $h(x)$. Here $c_{\mathfrak{m},\lambda},c_{\mathfrak{m},h_1} \in \Z$ and $\forall i \in \{1,2,\ldots,k\}, s_{\mathfrak{m},i} \in \Z$ and $ \forall \beta\in \F_{q^2}, r_{\mathfrak{m},\beta} \in \Z $ and $\forall u \in U_{\mathfrak{m},P}, c_{\mathfrak{m},u} \in \Z-\{0\}$.\\ \\
Recall that $h(x) = g(x) \prod_{i=1}^k g_i(x)^{a_i}$ is a factorization where $g_i(x)$ are distinct irreducible polynomials in $\F[x]$. For $i \in \{1,2,\ldots,k\}$, let $\mathcal{V}_i: \F_{q^2}(x) \longrightarrow \Z$ denote the valuation at $g_i(x)\F_{q^2}[x]$.\\ \\
If $\forall i \in \{1,2,\ldots,k\}$ and $\forall \beta \in \F_{q^2}$, $\mathcal{V}_i(P(x) - \beta) = 0$, then none of the factors of $h(x)$ can divide $N_{\mathfrak{m},P}$ and there is no need to look out for traps.\\ \\
If $\exists \beta \in \F_{q^2}$ and $\exists i \in \{1,2,\ldots,k\}$ such that $\mathcal{V}_i(P(x)-\beta) > 1$, then every $\mathfrak{m}$ that results in a relation involving $P(x)-\beta$ satisfies $\mathcal{V}_i(N_{\mathfrak{m},P})>1$. If $$\mathcal{V}_i(N_{\mathfrak{m},P}) = \sum_{\beta \in \F_{q^2}}\mathcal{V}_i(P(x)-\beta),$$ we can cancel the powers of $g_i(x) \mod g(x)$ and end up with $s_{\mathfrak{m},i}=0$. Else, the cancellation will result in $s_{\mathfrak{m},i} \neq 0$.\\ \\
Define $G_P$ to be $\left\{g_i(x)\ |\ \exists \mathfrak{m} \in \mathcal{P}_q: s_{\mathfrak{m},i} \neq 0\right\}$. In particular, $G_P$ is a subset of the set of irreducible factors of $h(x)$ that divide a translate of $P(x)$. Let $\hat{h}(x):=h(x)/(\prod_{g_i \in G_P} g_i(x))$. The relations that we obtain are for $\F_{\hat{h}}^\times$.\\ \\
If $N_{\mathfrak{m},P}(x)$ does factor over $\F_{q^2}[x]$ into a product of irreducible factors each of which is either of degree bounded by $w/2$ or divides $h(x)$, then from the relation obtained, form the relation vector $$R_{\mathfrak{m},P}:=\left(s_{\mathfrak{m},i}, r_{\mathfrak{m},\beta}\right)_{g_i(x) \in G_P, \beta \in \F_q^2} \in \Z^{|G_P|+q^2}$$ indexed by the elements in $G_P$ and $\F_{q^2}$. Let $M_P$ be the matrix consisting of $R_{\mathfrak{m},P}, \mathfrak{m} \in P_q$ as the rows where we only consider $\mathfrak{m}$ that resulted in a relation. Let $U_P$ denote the union of the sets $U_{\mathfrak{m},P}$ as $\mathfrak{m}$ ranges over elements in $\mathcal{P}_q$ that result in a relation.\\ \\
To apply results in \S~\ref{residue_class_rings} to a single step in the descent, we take $U$ to be the set of polynomials in $\F[x]$ of degree at most $w/2$, $\bar{U}$ to be $U_P$, $V$ to be $V_P$, $h(x)$ (of \S~\ref{residue_class_rings}) to be $\hat{h}(x)$ and $T$ to be the set of relations collected.\\ \\ 
Recall that $L$ equals $|\F_g^\times|$ divided by the largest $q^{2C}$-smooth factor of $|\F_g^\times|$. From the discussion in \S~\ref{poly_select}, we only need to perform the descent in the non smooth component $\F_g^\times[L]$.\\ \\
Our insistence on the modified polynomial selection results in $\F_h^\times(\ell)$ being cyclic for every prime $\ell$ dividing $L$. Hence, $\forall \ell \mid L, \F_h^\times(\ell) \cong \F_g^\times(\ell)$. Further, since $w>1$, $S \subseteq U$ and the image of $U$ in $\F_g^\times$ generates $\F_g^\times$. Thus the projection of $U$ in $\F_g^\times(\ell)$ generates $\F_g^\times(\ell)$ for all $\ell \mid L$ and the obstruction to descent does not arise leading to the following heuristic.
For every $\mathfrak{m} \in \mathcal{P}$, the degree of $N_{\mathfrak{m},P}(x)$ is bounded by $(1+D)w$. The probability that a random polynomial of degree at most $(1+D)w$ factors into irreducible factors of degree bounded by $w/2$ is around $1/(2(1+D))!$ which is a constant independent of $w$. Since there are $q(q^2+2)$ choices for $\mathfrak{m}$, if $N_{\mathfrak{m},P}(x)$ were to factor into irreducible polynomials of degree bounded by $w/2$ with a probability identical to that of a random polynomial of the same degree, then we expect to get at least $\Theta(q^3)$ relations. The number of columns in $M_P$ is bounded by $q^2+(q+D-m)/2$ and is likely to be close to $q^2$. The number of relations generated is likely to far exceed the number of columns in $M_P$ and thus $M_P$ is likely to have rank $q^2+|G_p|$ over $\mathbb{Q}$ and we arrive at the following heuristic.
\begin{heuristic}\label{heuristic_descent}
Let $T \subset \Z S$ denote the set of relations collected by a step in the Barbulescu-Gaudry-Joux-Thome descent algorithm starting from the polynomial $P(x)$ with the modified polynomial selection. Let $L$ denote $|\F_g^\times|$ divided by the largest $|\F|^C$-smooth factor of $|\F_g^\times|$. Then, for every prime $\ell$ dividing $L$, the $\Z/\ell\Z$ rank of $\Z T_V \otimes \Z/\ell\Z$ is $|V|$. 
\end{heuristic}
\noindent If heuristic \ref{heuristic_descent} is true, then at each step of the descent, we reduce the problem of descent from a polynomial $P(x)$ of degree $w$ to the problem of descent from a set of polynomials $V_P$ of degree at most $w/2$. A step in the descent can be performed in $q^{\mathcal{O}(1)}$ time by theorem \ref{theorem_gg}. Further, the size of $V_P$ is at most $\mathcal{O}(q^2w)$ \cite{bgjt}.  Since at each step we have at most $\mathcal{O}(q^2w)$ new descent steps involving polynomials of degree at most $w/2$, the total running time of the descent is $q^{\mathcal{O}(\log w)} = q^{\mathcal{O}(\log m)}$. the discussion in \S~\ref{poly_select}, the Pohlig-Hellman algorithm, theorem \ref{theorem_gg} and theorem \ref{dlog_factorbase_theorem} together imply the following theorem.
\begin{theorem}\label{dlog_descent_theorem}
Under heuristics \ref{heu1}, \ref{heuristic_factorbase} and \ref{heuristic_descent}, there is a deterministic algorithm that in time polynomial in $q^{\log(\deg(g))}$ solves the discrete logarithm problem in $\F_g^\times \cong \F_{q^{2m}}^\times$.
\end{theorem}
\noindent Under heuristics, we can thus find discrete logarithms in $\F_{p^n}^\times$ time polynomial in $p^{\log(n)}$ and $n^{\log(n)}$ and the algorithm is efficient (quasi polynomial) in small characteristic.\\ \\
If Heuristic \ref{heuristic_descent} fails for some $u(x)$ in the descent tree starting from a polynomial $P(x)$, then we may try again by taking a random power of $P(x)$ modulo $h(x)$.

\bibliographystyle{amsplain}

\end{document}